\documentclass[%
 aip,
 jmp,
 amsmath,amssymb,
preprint,%
]{revtex4-1}

\usepackage[dvips]{color}
\usepackage{graphicx}
\usepackage{dcolumn}
\usepackage{bm}
\usepackage{ulem}

\newcommand{\GaAs}{Al$_{0.33}$Ga$_{0.67}$As/GaAs}

\begin{document}
\title{\color{blue}{Anti-bunched photons from a lateral light-emitting diode}} %Title of paper
\author{Tommaso Lunghi}
\affiliation{Scuola Normale Superiore, Laboratorio NEST, I-56126 Pisa, Italy}
\author{Giorgio De Simoni}
\affiliation{Scuola Normale Superiore, Laboratorio NEST, I-56126 Pisa, Italy}
\author{Vincenzo Piazza}
\affiliation{Center for Nanotechnology Innovation @NEST, Istituto Italiano di Tecnologia, Piazza San Silvestro 12, 56127 Pisa}
\author{Christine A. Nicoll}
\affiliation{Cavendish Laboratory, University of Cambridge, Cambridge CB3 0HE, United Kingdom}
\author{Harvey E. Beere}
\affiliation{Cavendish Laboratory, University of Cambridge, Cambridge CB3 0HE, United Kingdom}
\author{David A. Ritchie}
\affiliation{Cavendish Laboratory, University of Cambridge, Cambridge CB3 0HE, United Kingdom}
\author{Fabio Beltram}
\affiliation{Scuola Normale Superiore, Laboratorio NEST, I-56126 Pisa, Italy}
%\date{\today}   

\begin{abstract}
We demonstrate anti-bunched emission from a lateral-light emitting diode. Sub-Poissonian emission statistic, with a g$^{(2)}$(0)=0.7, is achieved at cryogenic temperature in the pulsed low-current regime, by exploiting electron injection through shallow impurities located in the diode depletion region. Thanks to its simple fabrication scheme and to its modulation bandwidth in the GHz range, we believe our devices are an appealing substitute for highly-attenuated lasers in existing quantum-key-distribution systems. Our devices outperform strongly-attenuated lasers in terms of multi-photon emission events and can therefore lead to a significant security improvement in existing quantum key distribution systems. 
\end{abstract}

\maketitle

Quantum key distribution (QKD) makes it possible to achieve intrinsically-secure information sharing even through insecure classical communication channels\cite{Bennett1984}. QKD robustness was demonstrated both theoretically and experimentally, and today complete quantum cryptographic systems are commercially available\cite{Kollmitzer2010,Scarani2009}. Most QKD protocols rely on the assumption that each bit forming the encryption-decryption key is coded in the state of a quantum system, typically a single photon, which can be manipulated and transmitted by standard optical-telecommunication techniques.\\
In the literature several devices were proposed as ideal photon sources for QKD. Among them we would like to mention the electrically driven single-photon source\cite{Yuan2002} based on quantum dot which emits light strongly antibunched ($g^{(2)}(0)<0.13\pm 0.05$) and with high collection efficiency (34\%)\cite{Heindel2010},  but suffer from a slow and difficult fabrication procedure. To date most of the practical realizations are based on weak coherent pulses (WCPs)\cite{Brassard2000,Lutkenhaus2000} generated by strongly-attenuated lasers. WCPs are just an approximation to true single-photon pulses and consequently allow eavesdropping by photon-number splitting (PNS). In order to preserve transmission security with WCP sources, the average number of photons per pulse need be decreased. This leads to a reduced transmission rate and to a reduced achievable safe-transmission distance in lossy channels (or the implementation of PNS-robust techniques, e.~g.~decoy-states-based protocols\cite{Hwang2003}).

In this letter we present a lateral light-emitting diode (LLED), in which anti-bunched emission is achieved in the low-current pulsed-injection regime. This device architecture is based on a very simple fabrication and operation scheme \cite{Cecchini2003} and offers high modulation bandwidth in the GHz range \cite{Cecchini2004,Cecchini2005}. As a consequence we believe the present approach is very appealing as a substitute for highly-attenuated lasers that can be readily exploited in existing QKD systems.

Our devices were realized starting from a \textit{p}-type modulation doped \GaAs, heterostructure containing a 2-dimensional hole gas (2DHG) (the calculated charge density is $1.0\cdot 10^{11}$ cm$^{-2}$ at 4 K temperature) confined by a 20 nm-wide quantum well (QW). 
The heterostructure was processed into mesas and \textit{p}-type Au/Zn/Au annealed Ohmic contacts (5/55/100 nm) were thermally evaporated to provide electrical access to the 2DHG. 
The \textit{n}-type region of the LLEDs was fabricated by removing the \textit{p}-doping layer in a portion of the mesa by means of a wet-etching procedure and by evaporating a self-aligned annealed \textit{n}-type Ni/AuGe/Ni/Au (10/180/10/100 nm) Ohmic contact. The AuGe alloy provides donors to the host heterostructure generating an electron gas beneath the metal pad (Fig. 1(a)).

\begin{figure}[!ht]
\centering
\includegraphics[width=12cm]{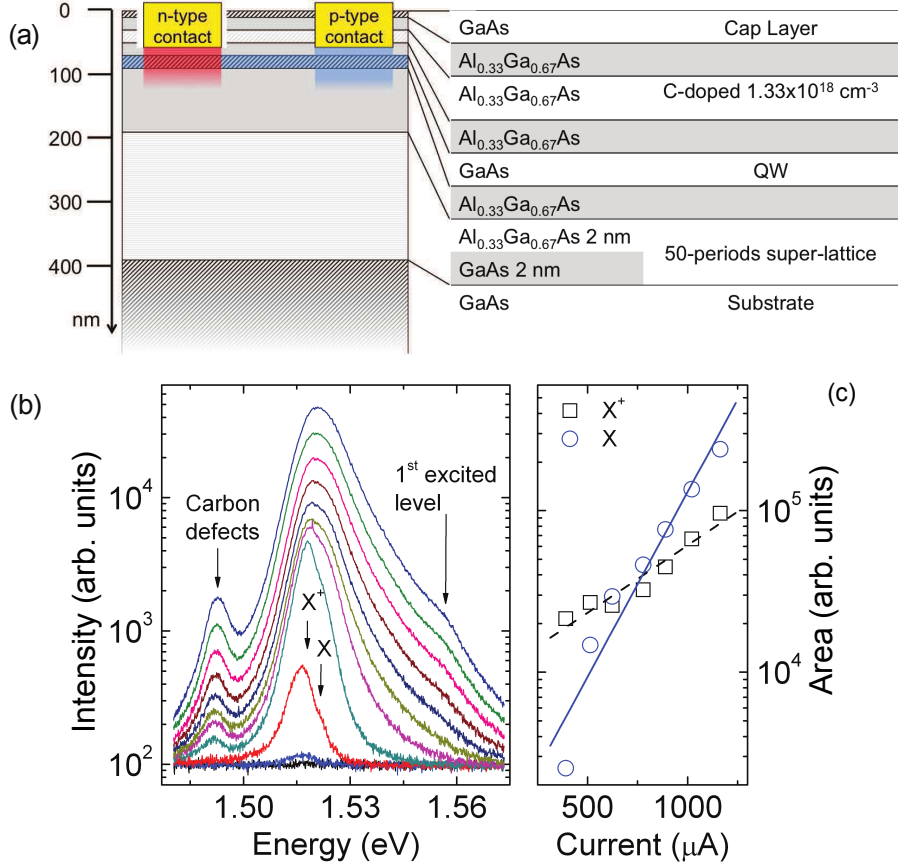}
\caption{(a) Fabrication scheme of the \textit{p-n} lateral junction. The yellow shapes represent the \textit{p}-type and \textit{n}-type contacts. The red and blue shaded areas respectively represent the n- and p- doped region induced by the annealing procedure. (b) Electroluminescence spectra collected increasing the forward bias from 1.45 V to 1.95 V in steps of 0.05 V at a temperature of 10 K. At lower injection current the emission, characterized by a double-peak structure, is originated by recombination of neutral excitons and positively-charged excitons. Increasing the injection current a third peak at lower energy appears originating from recombination of electron trapped in carbon defects within the GaAs. At higher injection current a fourth line appears due to recombination from the first excited sub-bands. (c) X, blue dots (X$^{+}$, black squares): areas of the emission peaks centered at 1.523 eV (1.518 eV) as a function of the injection current in logarithmic scale. The areas are derived by means of a best-fitting procedure with a sum of two Lorentzian. The linear fits, shown as dashed lines, yielded slope values of  A$_{X}=2.3 \pm 0.3$ (A$_{X^{+}}=0.8\pm 0.1 $). This is consistent with the expected faster increase of the peak area of the neutral excitons with respect to that of the positively charged exciton in a 2DHG\cite{Shields1995}.}
\label{fig:one}
\end{figure}

The actual formation of a lateral \textit{p-n} junction within the well at the interface between the 2DHG and the electron gas was confirmed by means of current-voltage characterization which showed the expected rectifying behavior with a conduction threshold ($\sim1.5$ V at 20 K) consistent with GaAs band gap. 
The electroluminescence (EL) signal was collected by a 0.68 numerical-aperture aspherical lens providing $\sim 1 \mu$m spatial resolution. EL spectra as a function of the junction forward bias ($V_f$) are reported in Fig.~\ref{fig:one}(b). They exhibit a double-peak structure originating from the recombination of neutral (X, $E_x=1.523$ eV) and charged (X$^+$, $E_{X^+}=1.518$ eV) excitons within the GaAs QW. These recombination mechanisms were identified\cite{Shields1995} by studying the evolution of peak areas as a function of the injection current (Fig.~\ref{fig:one}(c)). Additionally, when $V_f \gtrsim 1.6$ V, a shoulder is also visible at 1.558 eV, which we attribute to holes recombining with electrons populating the first-excited subband. This is confirmed by a self-consistent Poisson-Schr\"odinger calculation yielding an energy difference between the ground and the first-excited subband equal to 32 meV. A peak due to electron recombination at carbon defects is also present. Moreover polarization resolved electroluminscence measurements revealed that emitted photons are both clock- and counter-clock-wise circularly polarized in the plane of the QW, consistently with recombination mechanism in a GaAs QW.

The second-order correlation function $g^{(2)}(\tau)$ of the emitted photons was measured by means of a time-correlated single-photon-counting (TCSPC) \cite{Brown1956} setup which directly produces the correlation count histogram, $K(\tau)$, as a function of the time delay $\tau$ between the arrival times of two successive photons on two different detectors. The LLEDs were biased with periodic -5 V pulses added to a dc voltage ($V_{dc}$). Pulses were applied to the $n$-type contact while keeping the $p$-type contact grounded. Several combinations of $V_{dc}$, pulse widths ($t_p$) and periodicities ($T_p$) were explored. The rise-fall time of our pulse generator is of the order of 2 ns, so that when the nominal $t_p$ was set to a value $\lesssim 5$ ns triangular pulses were obtained. Taking into account that electron injection into the 2DHG occurs only when the applied voltage ($V_{dc}$ plus the pulse) exceeds the diode threshold, the peculiar shape of our pulses allowed us to drive the LLEDs in the sub-ns time scale. 

\begin{figure}[!ht]
\centering
\includegraphics[width=12cm]{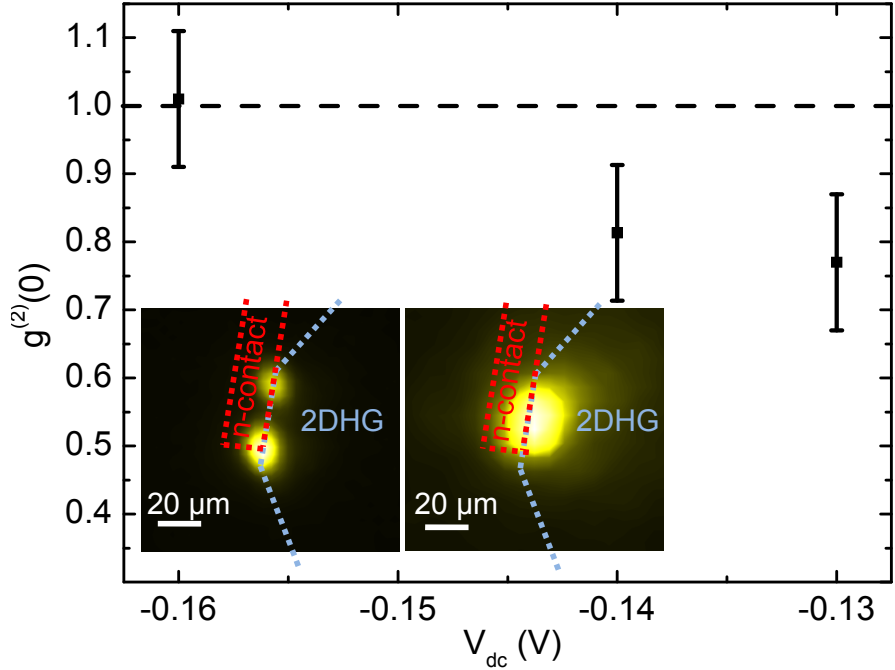}
\caption{g$^{(2)}(0)$ as a function of (V$_{dc}$). The \textit{p}-type contact was grounded. The value of g$^{(2)}(0)$ is derived from the ratio between the correlation counts K at $\tau$ =0 and the average among the correlation counts of the peaks with $\tau \neq 0$. Lowering the injection current the value of g$^{(2)}(0)$ decreases down to $0.8 \pm 0.1$ demonstrating anti-bunched emission. Insets: Spatially resolved electroluminescence measurements of the junction region in the spot-emission and in the high-injection regimes at a temperature of 4 K. The scale of the light intensity is in arbitrary units. Emission maps were recorded by focusing emitted light on a CCD camera, through a 10X microscope objective. The resolution of this setup is much lower than that obtained with the 0.68-NA aperture lens. Temperature was 4 K. The edges of the $n$-contact (red line) and of the 2DHG (blue line) are also reported. (a): Spot-emission regime. The device was forward biased with $V_{dc}=-0.96$  V plus voltage pulses with width and periodicity of 5 ns and 100 ns respectively. The scale of the light intensity is in arbitrary units. (b): High-injection regime. The device was forward biased with $V_{dc}=-3$ V.}
\label{fig:two}
\end{figure}

Figure \ref{fig:two} reports the measured $g^{(2)}(0)$ as function of $V_{dc}$, at $t_p = 3$ ns and  $T_p = 80$ ns. 
At $V_{dc} = -0.16$ V, i.e. in the high-current-injection regime, we found $g^{(2)}(0) = 1$ within error, consistently with the value expected for a Poissonian source. Reduction of the current by increasing $V_{dc}$ led to a suppression of $g^{(2)}(0)$, down to a value of $\sim 0.75$ at $V_{dc} = -0.13$ V.
In the low-current-injection regime, our LLEDs therefore behave as sub-Poissonian sources and hint at the possibility of true single-photon emission\cite{Davidovich1996}.

Insight into the mechanism leading to anti-bunched emission was provided by the analysis of the spatial distribution of the emitted light. Spatially-resolved EL measurements [inset (a) of \ref{fig:two}] show that, when LLEDs are biased just above conduction threshold, light emission occurs from localized spots in the junction region whose size is not larger than the spatial resolution of our optical setup ($\sim$1~ $\mu$m). This behavior is explained by the picture of electrons being injected through a series of shallow impurities (most probably Ge atoms originating from the Ohmic contact fabrication) located in the diode depletion region, whose energy is just below the conduction band bottom (Ge impurity binding energy is -5.88 meV with respect to the conduction band bottom\cite{Yu1996}). At cryogenic temperatures, the occupation number of impurity states is 0 or 1, due to the charging energy, which forbid more than one electron to simultaneously occupy the same impurity state. For these reason, at low current injection, \textit{i. e.} when the electrochemical potential of the $n$-section of the diode ($\mu_n$) is aligned with the impurity level energy, electrons can be injected one by one into the impurity centers, where radiative recombination takes place. Evidently, in these regime, light emission is, as observed, spatially localized (spot emission regime) and it is expected to be anti-bunched. A further increase of $\mu_n$ leads to electrons being directly injected into the conduction band in the $p$-section of the device. In this regime single-electron injection is not expected to occur: spot-emission regime is suppressed and luminescence comes from the whole region surrounding the n-contact (inset (b) of \ref{fig:two}). Considering the nature of the injection mechanism, we believe that the working temperature range of the device is determined by the Coulomb repulsion energy scale, which is typically $\sim$1 K and below. Temperature characterization of photon emission statistics confirmed this hypotheses, indeed anti-bunched light was not observed above 4K, which constitutes an experimental upper-bound for the device operation temperature.
 
In our experiments, we did not observe the complete vanishing of  $g^{(2)}(0)$. In fact the minimum observed value of $g^{(2)}(0) \sim 0.75$ can only be reduced to an estimated true value of $\sim 0.73$ by taking into account the dark counts of our detectors (25 s$^{-1}$). This demonstrates that multi-photon emission still occurs even if suppressed. A likely origin of this multi-photon emission may be found in the amplitude-jitter of the voltage source. In fact, if we define a single- ($V_{single}$) and a multi-electron ($V_{multi}$) injection threshold, when the forward bias overcomes $V_{multi}$ two or more electrons can be injected at a time, giving rise to multi-photon emissions and correlation counts at zero time delay. On the contrary, when $V_{single}< V_f(t) < V_{multi}$, only one electron at a time can be injected. Emission of a single-photon per pulse from one spot can be expected if in each excitation period $V_{single} < V_f(t) < V_{multi}$ for a time interval shorter than the electron radiative lifetime ($\tau_e \sim 1$ ns). The width of the peak at zero time delay is then given by the sum of $\tau_e$ and the time interval during which $V_f(t)  > V_{multi}$, while the width of the other peaks is the sum of $\tau_e$ and the time interval during which $V_f(t)  > V_{single}$. 

\begin{figure}[!ht]
\centering
\includegraphics[width=12cm]{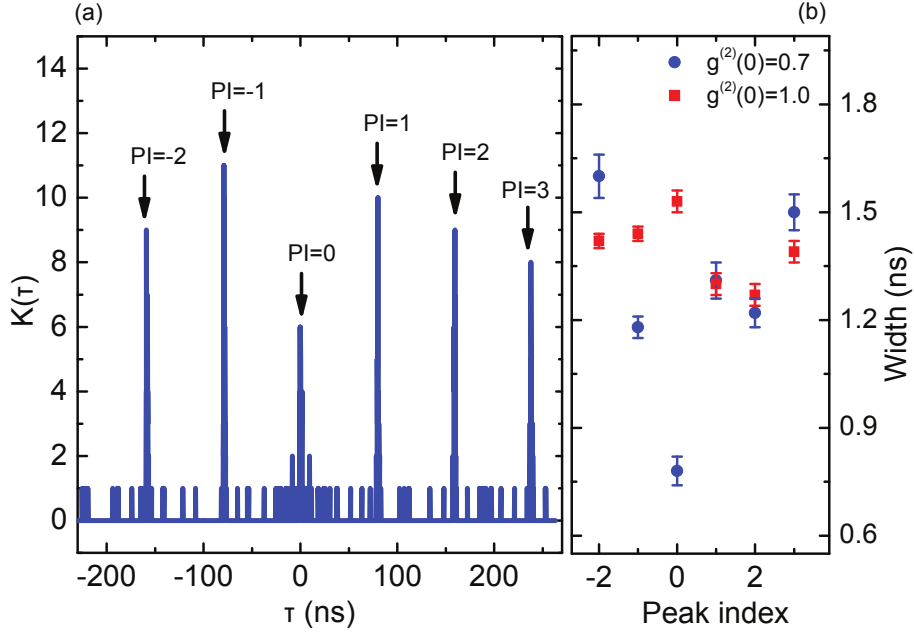}
\caption{(a) Correlation counts (K) as a function of the delay time ($\tau$). The LLED was biased with pulses of 3 ns of width, with a repetition period of 80 ns and V$_{dc}$=-0.86 V. The pulse amplitude was reduced by a 8-dB attenuator lowering the amplitude-jitter. After a collection time of 12 hours we measured about 80 correlation counts at each peak. The measurement shows equally-spaced peaks, indicated by the arrows and labeled with the peak index PI. A suppressed 0-time delay peak  proves the anti-bunched emission, as confirmed by the calculated value of $g^{(2)}(0)= 0.7 \pm 0.1$. (b) Width of the peaks as a function of the peak index when the LLED emits Poissonian light ($g^{(2)}(0)= 1.0 \pm 0.1$, red squares) and in antibunching regime ($g^{(2)}(0)= 0.7 \pm 0.1$, blue dots).}
\label{fig:three}
\end{figure}

To test this hypothesis we attenuated the pulse signal by 8 dB to reduce the amplitude jitter. In order to compensate for the lower excitation signal, we increased (in absolute value) $V_{dc}$. Figure \ref{fig:three}(a) shows the second-order correlation function collected in these conditions. The total number of coincidences K$_i$ for each peak $i$ was obtained by the integral of peaks area, yielding K$_0\sim$50 against the average value of $\sim$75 for the other peaks, which have a constant K$_{i\neq 0}$ value, within the error bar. Although the collection efficiency of present devices is quite low (collection time was $\sim$12 hours), it can be improved by embedding the LLED in a vertical cavity\cite{Bennett2005,Strauf2007} that is the standard technique to implement QD-based single photon sources. The extracted $g^{(2)}(0)$ for this data set was the lowest we could obtain and is 0.7$\pm$0.1. The analysis of the width of the peaks of $K(\tau)$ is reported in Fig. ~\ref{fig:three}(b) for the datasets collected at $V_{dc} = -0.16$ V and $V_{dc} = -0.13$ V (corresponding to $g^{(2)}(0)=  1.0 \pm 0.1$   (red squares) and $g^{(2)}(0)= 0.7 \pm 0.1$ (blue circles)). A reduction of the width of the peak at zero time delay for the second dataset is evident.

In conclusion, we demonstrated anti-bunched emission from a lateral \textit{p-n} junction by means of second-order-correlation-function measurements yielding a measured minimum value of g$^{2}(0)$=0.7$\pm$0.1. 
The width difference between the peak at zero-time delay and the peaks at different delays when the LLED emits in anti-bunching regime confirms the generation mechanism of the anti-bunched light proposed in this letter. 

%\bibliography{all,all2}

\begin{thebibliography}{17}%
\makeatletter
\providecommand \@ifxundefined [1]{%
 \@ifx{#1\undefined}
}%
\providecommand \@ifnum [1]{%
 \ifnum #1\expandafter \@firstoftwo
 \else \expandafter \@secondoftwo
 \fi
}%
\providecommand \@ifx [1]{%
 \ifx #1\expandafter \@firstoftwo
 \else \expandafter \@secondoftwo
 \fi
}%
\providecommand \natexlab [1]{#1}%
\providecommand \enquote  [1]{``#1''}%
\providecommand \bibnamefont  [1]{#1}%
\providecommand \bibfnamefont [1]{#1}%
\providecommand \citenamefont [1]{#1}%
\providecommand \href@noop [0]{\@secondoftwo}%
\providecommand \href [0]{\begingroup \@sanitize@url \@href}%
\providecommand \@href[1]{\@@startlink{#1}\@@href}%
\providecommand \@@href[1]{\endgroup#1\@@endlink}%
\providecommand \@sanitize@url [0]{\catcode `\\12\catcode `\$12\catcode
  `\&12\catcode `\#12\catcode `\^12\catcode `\_12\catcode `\%12\relax}%
\providecommand \@@startlink[1]{}%
\providecommand \@@endlink[0]{}%
\providecommand \url  [0]{\begingroup\@sanitize@url \@url }%
\providecommand \@url [1]{\endgroup\@href {#1}{\urlprefix }}%
\providecommand \urlprefix  [0]{URL }%
\providecommand \Eprint [0]{\href }%
\providecommand \doibase [0]{http://dx.doi.org/}%
\providecommand \selectlanguage [0]{\@gobble}%
\providecommand \bibinfo  [0]{\@secondoftwo}%
\providecommand \bibfield  [0]{\@secondoftwo}%
\providecommand \translation [1]{[#1]}%
\providecommand \BibitemOpen [0]{}%
\providecommand \bibitemStop [0]{}%
\providecommand \bibitemNoStop [0]{.\EOS\space}%
\providecommand \EOS [0]{\spacefactor3000\relax}%
\providecommand \BibitemShut  [1]{\csname bibitem#1\endcsname}%
\let\auto@bib@innerbib\@empty
%</preamble>
\bibitem [{\citenamefont {Bennett}\ and\ \citenamefont
  {Brassard}(1984)}]{Bennett1984}%
  \BibitemOpen
  \bibfield  {author} {\bibinfo {author} {\bibfnamefont {C.}~\bibnamefont
  {Bennett}}\ and\ \bibinfo {author} {\bibfnamefont {G.}~\bibnamefont
  {Brassard}},\ }\href@noop
  {}  {\bibinfo {booktitle} {Proc. of IEEE Int. Conf.
  on Computers, Systems and Signal Processing}}, \bibinfo {volume}
  {175}\ (\bibinfo {year}
  {1984})\BibitemShut {NoStop}%
\bibitem [{\citenamefont {Kollmitzer}\ and\ \citenamefont
  {Pivk}(2010)}]{Kollmitzer2010}%
  \BibitemOpen
  \bibfield  {author} {\bibinfo {author} {\bibfnamefont {C.}~\bibnamefont
  {Kollmitzer}}\ and\ \bibinfo {author} {\bibfnamefont {M.}~\bibnamefont
  {Pivk}},\ }\href@noop {} {\bibinfo {title} {Applied Quantum
  Cryptography}}\ (\bibinfo  {publisher} {Springer-Verlag},\ \bibinfo {year}
  {2010})\ \BibitemShut {NoStop}%
\bibitem [{\citenamefont {Scarani}\ \emph {et~al.}(2009)\citenamefont
  {Scarani}, \citenamefont {Bechmann-Pasquinucci}, \citenamefont {Cerf},
  \citenamefont {Du\v{s}ek}, \citenamefont {L\"{u}tkenhaus},\ and\
  \citenamefont {Peev}}]{Scarani2009}%
  \BibitemOpen
  \bibfield  {author} {\bibinfo {author} {\bibfnamefont {V.}~\bibnamefont
  {Scarani}}, \bibinfo {author} {\bibfnamefont {H.}~\bibnamefont
  {Bechmann-Pasquinucci}}, \bibinfo {author} {\bibfnamefont {N.}~\bibnamefont
  {Cerf}}, \bibinfo {author} {\bibfnamefont {M.}~\bibnamefont {Du\v{s}ek}},
  \bibinfo {author} {\bibfnamefont {N.}~\bibnamefont {L\"{u}tkenhaus}}, \ and\
  \bibinfo {author} {\bibfnamefont {M.}~\bibnamefont {Peev}},\ }\href@noop {}
  {\bibfield  {journal} {\bibinfo  {journal} {Rev. of Mod. Phys.}\ }\textbf
  {\bibinfo {volume} {81}},\ \bibinfo {pages} {1301} (\bibinfo {year}
  {2009})}\BibitemShut {NoStop}%
\bibitem [{\citenamefont {Yuan}\ \emph {et~al.}(2002)\citenamefont {Yuan},
  \citenamefont {Kardynal}, \citenamefont {Stevenson}, \citenamefont {Shields},
  \citenamefont {Lobo}, \citenamefont {Cooper}, \citenamefont {Beattie},
  \citenamefont {Ritchie},\ and\ \citenamefont {Pepper}}]{Yuan2002}%
  \BibitemOpen
  \bibfield  {author} {\bibinfo {author} {\bibfnamefont {Z.}~\bibnamefont
  {Yuan}}, \bibinfo {author} {\bibfnamefont {B.~E.}\ \bibnamefont {Kardynal}},
  \bibinfo {author} {\bibfnamefont {R.~M.}\ \bibnamefont {Stevenson}}, \bibinfo
  {author} {\bibfnamefont {A.~J.}\ \bibnamefont {Shields}}, \bibinfo {author}
  {\bibfnamefont {C.~J.}\ \bibnamefont {Lobo}}, \bibinfo {author}
  {\bibfnamefont {K.}~\bibnamefont {Cooper}}, \bibinfo {author} {\bibfnamefont
  {N.~S.}\ \bibnamefont {Beattie}}, \bibinfo {author} {\bibfnamefont {D.~a.}\
  \bibnamefont {Ritchie}}, \ and\ \bibinfo {author} {\bibfnamefont
  {M.}~\bibnamefont {Pepper}},\ }\href@noop {} {\bibfield  {journal} {\bibinfo
  {journal} {Science}\ }\textbf {\bibinfo {volume} {295}} (\bibinfo {year}
  {2002})}\BibitemShut {NoStop}%
\bibitem [{\citenamefont {Heindel}\ \emph {et~al.}(2010)\citenamefont
  {Heindel}, \citenamefont {Schneider}, \citenamefont {Lermer}, \citenamefont
  {Kwon}, \citenamefont {Braun}, \citenamefont {Reitzenstein}, \citenamefont
  {Hofling}, \citenamefont {Kamp},\ and\ \citenamefont
  {Forchel}}]{Heindel2010}%
  \BibitemOpen
  \bibfield  {author} {\bibinfo {author} {\bibfnamefont {T.}~\bibnamefont
  {Heindel}}, \bibinfo {author} {\bibfnamefont {C.}~\bibnamefont {Schneider}},
  \bibinfo {author} {\bibfnamefont {M.}~\bibnamefont {Lermer}}, \bibinfo
  {author} {\bibfnamefont {S.~H.}\ \bibnamefont {Kwon}}, \bibinfo {author}
  {\bibfnamefont {T.}~\bibnamefont {Braun}}, \bibinfo {author} {\bibfnamefont
  {S.}~\bibnamefont {Reitzenstein}}, \bibinfo {author} {\bibfnamefont
  {S.}~\bibnamefont {Hofling}}, \bibinfo {author} {\bibfnamefont
  {M.}~\bibnamefont {Kamp}}, \ and\ \bibinfo {author} {\bibfnamefont
  {A.}~\bibnamefont {Forchel}},\ }\href@noop {} {\bibfield  {journal} {\bibinfo
   {journal} {Appl. Phys. Lett.}\ }\textbf {\bibinfo {volume} {96}},\ \bibinfo
  {pages} {011107} (\bibinfo {year} {2010})}\BibitemShut {NoStop}%
\bibitem [{\citenamefont {Brassard}\ \emph {et~al.}(2000)\citenamefont
  {Brassard}, \citenamefont {Lutkenhaus}, \citenamefont {Mor},\ and\
  \citenamefont {Sanders}}]{Brassard2000}%
  \BibitemOpen
  \bibfield  {author} {\bibinfo {author} {\bibfnamefont {G.}~\bibnamefont
  {Brassard}}, \bibinfo {author} {\bibfnamefont {N.}~\bibnamefont
  {Lutkenhaus}}, \bibinfo {author} {\bibfnamefont {T.}~\bibnamefont {Mor}}, \
  and\ \bibinfo {author} {\bibfnamefont {B.~C.}\ \bibnamefont {Sanders}},\
  }\href@noop {} {\bibfield  {journal} {\bibinfo  {journal} {Phys. Rev. Lett.}\
  }\textbf {\bibinfo {volume} {85}},\ \bibinfo {pages} {1330} (\bibinfo {year}
  {2000})}\BibitemShut {NoStop}%
\bibitem [{\citenamefont {L\"{u}tkenhaus}(2000)}]{Lutkenhaus2000}%
  \BibitemOpen
  \bibfield  {author} {\bibinfo {author} {\bibfnamefont {N.}~\bibnamefont
  {L\"{u}tkenhaus}},\ }\href@noop {} {\bibfield  {journal} {\bibinfo  {journal}
  {Phys. Rev. A}\ }\textbf {\bibinfo {volume} {61}},\ \bibinfo {pages} {1}
  (\bibinfo {year} {2000})}\BibitemShut {NoStop}%
\bibitem [{\citenamefont {Hwang}(2003)}]{Hwang2003}%
  \BibitemOpen
  \bibfield  {author} {\bibinfo {author} {\bibfnamefont {W.}~\bibnamefont
  {Hwang}},\ }\href@noop {} {\bibfield  {journal} {\bibinfo  {journal} {Phys.
  Rev. Lett.}\ }\textbf {\bibinfo {volume} {91}},\ \bibinfo {pages} {1}
  (\bibinfo {year} {2003})}\BibitemShut {NoStop}%
\bibitem [{\citenamefont {Cecchini}\ \emph {et~al.}(2003)\citenamefont
  {Cecchini}, \citenamefont {Piazza}, \citenamefont {Beltram}, \citenamefont
  {Lazzarino}, \citenamefont {Ward}, \citenamefont {Shields}, \citenamefont
  {Beere},\ and\ \citenamefont {Ritchie}}]{Cecchini2003}%
  \BibitemOpen
  \bibfield  {author} {\bibinfo {author} {\bibfnamefont {M.}~\bibnamefont
  {Cecchini}}, \bibinfo {author} {\bibfnamefont {V.}~\bibnamefont {Piazza}},
  \bibinfo {author} {\bibfnamefont {F.}~\bibnamefont {Beltram}}, \bibinfo
  {author} {\bibfnamefont {M.}~\bibnamefont {Lazzarino}}, \bibinfo {author}
  {\bibfnamefont {M.~B.}\ \bibnamefont {Ward}}, \bibinfo {author}
  {\bibfnamefont {A.~J.}\ \bibnamefont {Shields}}, \bibinfo {author}
  {\bibfnamefont {H.~E.}\ \bibnamefont {Beere}}, \ and\ \bibinfo {author}
  {\bibfnamefont {D.~A.}\ \bibnamefont {Ritchie}},\ }\href@noop {} {\bibfield
  {journal} {\bibinfo  {journal} {Appl. Phys. Let.}\ }\textbf {\bibinfo
  {volume} {82}},\ \bibinfo {pages} {636} (\bibinfo {year} {2003})}\BibitemShut
  {NoStop}%
\bibitem [{\citenamefont {Cecchini}\ \emph {et~al.}(2004)\citenamefont
  {Cecchini}, \citenamefont {{De Simoni}}, \citenamefont {Piazza},
  \citenamefont {Beltram}, \citenamefont {Beere},\ and\ \citenamefont
  {Ritchie}}]{Cecchini2004}%
  \BibitemOpen
  \bibfield  {author} {\bibinfo {author} {\bibfnamefont {M.}~\bibnamefont
  {Cecchini}}, \bibinfo {author} {\bibfnamefont {G.}~\bibnamefont {{De
  Simoni}}}, \bibinfo {author} {\bibfnamefont {V.}~\bibnamefont {Piazza}},
  \bibinfo {author} {\bibfnamefont {F.}~\bibnamefont {Beltram}}, \bibinfo
  {author} {\bibfnamefont {H.~E.}\ \bibnamefont {Beere}}, \ and\ \bibinfo
  {author} {\bibfnamefont {D.~A.}\ \bibnamefont {Ritchie}},\ }\href@noop {}
  {\bibfield  {journal} {\bibinfo  {journal} {Appl. Phys. Lett.}\ }\textbf
  {\bibinfo {volume} {85}},\ \bibinfo {pages} {3020} (\bibinfo {year}
  {2004})}\BibitemShut {NoStop}%
\bibitem [{\citenamefont {Cecchini}\ \emph {et~al.}(2005)\citenamefont
  {Cecchini}, \citenamefont {Piazza}, \citenamefont {Beltram}, \citenamefont
  {Gevaux}, \citenamefont {Ward}, \citenamefont {Shields}, \citenamefont
  {Beere},\ and\ \citenamefont {Ritchie}}]{Cecchini2005}%
  \BibitemOpen
  \bibfield  {author} {\bibinfo {author} {\bibfnamefont {M.}~\bibnamefont
  {Cecchini}}, \bibinfo {author} {\bibfnamefont {V.}~\bibnamefont {Piazza}},
  \bibinfo {author} {\bibfnamefont {F.}~\bibnamefont {Beltram}}, \bibinfo
  {author} {\bibfnamefont {D.~G.}\ \bibnamefont {Gevaux}}, \bibinfo {author}
  {\bibfnamefont {M.~B.}\ \bibnamefont {Ward}}, \bibinfo {author}
  {\bibfnamefont {A.~J.}\ \bibnamefont {Shields}}, \bibinfo {author}
  {\bibfnamefont {H.~E.}\ \bibnamefont {Beere}}, \ and\ \bibinfo {author}
  {\bibfnamefont {D.~A.}\ \bibnamefont {Ritchie}},\ }\href@noop {} {\bibfield
  {journal} {\bibinfo  {journal} {Appl. Phys. Lett.}\ }\textbf {\bibinfo
  {volume} {86}},\ \bibinfo {pages} {241107} (\bibinfo {year}
  {2005})}\BibitemShut {NoStop}%
\bibitem [{\citenamefont {Shields}\ \emph {et~al.}(1995)\citenamefont
  {Shields}, \citenamefont {Osborne}, \citenamefont {Simmons}, \citenamefont
  {Pepper},\ and\ \citenamefont {Ritchie}}]{Shields1995}%
  \BibitemOpen
  \bibfield  {author} {\bibinfo {author} {\bibfnamefont {A.~J.}\ \bibnamefont
  {Shields}}, \bibinfo {author} {\bibfnamefont {J.~L.}\ \bibnamefont
  {Osborne}}, \bibinfo {author} {\bibfnamefont {M.~Y.}\ \bibnamefont
  {Simmons}}, \bibinfo {author} {\bibfnamefont {M.}~\bibnamefont {Pepper}}, \
  and\ \bibinfo {author} {\bibfnamefont {D.~A.}\ \bibnamefont {Ritchie}},\
  }\href@noop {} {\bibfield  {journal} {\bibinfo  {journal} {Phys. Rev. B}\
  }\textbf {\bibinfo {volume} {52}},\ \bibinfo {pages} {R5523} (\bibinfo {year}
  {1995})}\BibitemShut {NoStop}%
\bibitem [{\citenamefont {Brown}\ and\ \citenamefont
  {Twiss}(1956)}]{Brown1956}%
  \BibitemOpen
  \bibfield  {author} {\bibinfo {author} {\bibfnamefont {R.}~\bibnamefont
  {Brown}}\ and\ \bibinfo {author} {\bibfnamefont {R.~Q.}\ \bibnamefont
  {Twiss}},\ }\href@noop {} {\bibfield  {journal} {\bibinfo  {journal}
  {Nature}\ }\textbf {\bibinfo {volume} {177}},\ \bibinfo {pages} {27}
  (\bibinfo {year} {1956})}\BibitemShut {NoStop}%
\bibitem [{\citenamefont {Davidovich}(1996)}]{Davidovich1996}%
  \BibitemOpen
  \bibfield  {author} {\bibinfo {author} {\bibfnamefont {L.}~\bibnamefont
  {Davidovich}},\ }\href@noop {} {\bibfield  {journal} {\bibinfo  {journal}
  {Rev. Mod. Phys.}\ }\textbf {\bibinfo {volume} {68}},\ \bibinfo {pages} {127}
  (\bibinfo {year} {1996})}\BibitemShut {NoStop}%
\bibitem [{\citenamefont {Yu}\ and\ \citenamefont {Cardona}(1996)}]{Yu1996}%
  \BibitemOpen
  \bibfield  {author} {\bibinfo {author} {\bibfnamefont {P.}~\bibnamefont
  {Yu}}\ and\ \bibinfo {author} {\bibfnamefont {M.}~\bibnamefont {Cardona}},\
  }\href@noop {}  {\bibinfo {title} {{Fundamentals of semiconductors:
  physics and materials properties}}}\ (\bibinfo  {publisher}
  {Springer-Verlag},\ \bibinfo {year}
  {1996})\BibitemShut {NoStop}%
\bibitem [{\citenamefont {Bennett}\ \emph {et~al.}(2005)\citenamefont
  {Bennett}, \citenamefont {Unitt}, \citenamefont {See}, \citenamefont
  {Shields}, \citenamefont {Atkinson}, \citenamefont {Cooper},\ and\
  \citenamefont {Ritchie}}]{Bennett2005}%
  \BibitemOpen
  \bibfield  {author} {\bibinfo {author} {\bibfnamefont {A.~J.}\ \bibnamefont
  {Bennett}}, \bibinfo {author} {\bibfnamefont {D.~C.}\ \bibnamefont {Unitt}},
  \bibinfo {author} {\bibfnamefont {P.}~\bibnamefont {See}}, \bibinfo {author}
  {\bibfnamefont {A.~J.}\ \bibnamefont {Shields}}, \bibinfo {author}
  {\bibfnamefont {P.}~\bibnamefont {Atkinson}}, \bibinfo {author}
  {\bibfnamefont {K.}~\bibnamefont {Cooper}}, \ and\ \bibinfo {author}
  {\bibfnamefont {D.~A.}\ \bibnamefont {Ritchie}},\ }\href@noop {} {\bibfield
  {journal} {\bibinfo  {journal} {Appl. Phys. Lett.}\ }\textbf {\bibinfo
  {volume} {86}},\ \bibinfo {pages} {181102} (\bibinfo {year}
  {2005})}\BibitemShut {NoStop}%
\bibitem [{\citenamefont {Strauf}\ \emph {et~al.}(2007)\citenamefont {Strauf},
  \citenamefont {Stoltz}, \citenamefont {Rakher}, \citenamefont {Coldren},
  \citenamefont {Petroff},\ and\ \citenamefont {Bouwmeester}}]{Strauf2007}%
  \BibitemOpen
  \bibfield  {author} {\bibinfo {author} {\bibfnamefont {S.}~\bibnamefont
  {Strauf}}, \bibinfo {author} {\bibfnamefont {N.~G.}\ \bibnamefont {Stoltz}},
  \bibinfo {author} {\bibfnamefont {M.~T.}\ \bibnamefont {Rakher}}, \bibinfo
  {author} {\bibfnamefont {L.~A.}\ \bibnamefont {Coldren}}, \bibinfo {author}
  {\bibfnamefont {P.~M.}\ \bibnamefont {Petroff}}, \ and\ \bibinfo {author}
  {\bibfnamefont {D.}~\bibnamefont {Bouwmeester}},\ }\href@noop {} {\bibfield
  {journal} {\bibinfo  {journal} {Nature Photonics}\ }\textbf {\bibinfo
  {volume} {1}},\ \bibinfo {pages} {704} (\bibinfo {year} {2007})}\BibitemShut
  {NoStop}%
\end{thebibliography}
%

\end{document}